\documentclass[prb,aps,nobibnotes,twocolumn,doublespace,twocolumngrid,superbib,showpacs]{revtex4}

\usepackage{graphicx}
\usepackage{amsfonts}
\usepackage{amsmath}
\usepackage{bm}
\usepackage{alltt}
\usepackage{dcolumn} 
\usepackage{graphicx}

\begin{document}

\title{Linear scaling computation of the Fock matrix.  VIII. \\  
       Periodic boundaries for exact exchange at the $\Gamma$-point.$^1$ }

\author{C. J. Tymczak$^2$}
\affiliation{ Theoretical Division, Los Alamos National Laboratory, Los Alamos, New Mexico 87545 }
\author{Val{\'e}ry T.~Weber}
\affiliation{Dept.~of Chemistry, University of Fribourg, 1700 Fribourg, Switzerland}
\author{Eric Schwegler}
\affiliation{Lawrence Livermore National Laboratory, Livermore, California 94551}
\author{Matt Challacombe}
\affiliation{Theoretical Division, Los Alamos National Laboratory, Los Alamos, New Mexico 87545 }

\date{\today}
\begin{abstract}
A translationally invariant formulation of the Hartree-Fock (HF) $\Gamma$-point approximation
is presented.   This formulation is achieved through introduction of  the  Minimum Image Convention (MIC) at 
the level of primitive two-electron integrals, and implemented in a periodic version of the 
ONX algorithm [J.~Chem.~Phys, {\bf 106} 9708 (1997)] for linear scaling computation of the
exchange matrix. Convergence of the HF-MIC $\Gamma$-point model to the HF ${\bf k}$-space limit 
is demonstrated for fully periodic magnesium oxide, ice and diamond.  Computation of the diamond
lattice constant using the HF-MIC model together with the hybrid PBE0 density functional 
[Theochem, {\bf 493} 145 (1999)] yields $a_0=3.569$\AA~ with the 6-21G* basis set and a 
$3\times3\times3$ supercell.  Linear scaling computation of the HF-MIC exchange matrix is demonstrated 
for diamond and ice in the condensed phase. 
\end{abstract}


\maketitle

\footnotetext[1]{\tt Preprint LA-UR 03-9043}
\footnotetext[2]{\tt tymczak@lanl.gov}

\section{INTRODUCTION}

In a preceding article \cite{CTymczak04A},  methods were introduced for constructing 
the periodic $\Gamma$-point Coulomb and Exchange-Correlation matrix that 
achieve  a cost  scaling only linearly with system size, $N$.  In this paper, an ${\cal O}(N)$
algorithm is presented for computation of the periodic Hartree-Fock (HF) exchange matrix at the 
$\Gamma$-point.   The Hartree-Fock approximation is often a fast, first 
approximation and also a starting point for correlated ``wavefunction'' methods.  
Also, hybrid Hartree-Fock/Density Functional Theory (HF/DFT) model chemistries are an important next 
step in accuracy beyond the Generalized Gradient Approximation \cite{Gill92,Becke93,VBarone96,CAdamo99}.
Together with linear scaling methods for computing the density matrix \cite{ANiklasson02A,ANiklasson03}, these 
advances provide a framework for the application of both HF and HF/DFT models to large condensed 
phase systems, surfaces and wires.   

To date, condensed phase HF and HF/DFT calculations have been carried out almost 
exclusively with the {\em ab initio} solid state program {\sc CRYSTAL} \cite{RDovesi00}, 
which employs conventional  algorithms for evaluation of the 
HF exchange matrix.  The {\sc CRYSTAL} program employs methods for ${\bf k}$-space integration, 
which generates wavefunctions with the correct translational symmetries.   While the  
$\Gamma$-point approximation forgoes ${\bf k}$-space integration, it should recover the correct 
symmetries in the limit of a large super cell.  With a more tractable formulation and asymptotic 
correctness in the limit of large systems,  the $\Gamma$-point  approximation would seem an 
ideal basis for linear scaling exchange algorithms.   However in preliminary studies, we found 
that the naive HF $\Gamma$-point approximation converges to position dependent values, and that 
even in the large super cell limit,  values different from the ${\bf k}$-space integration limit 
are approached.

In this paper we develop a translationally invariant definition of $\Gamma$-point 
Hartree-Fock exchange, which correctly approaches the ${\bf k}$-space integration value
in the limit of a large super-cell.  This is accomplished through introducing a Minimum 
Image Convention (MIC) into the exchange kernel at the level of primitive two-electron integrals,  
as described in Section \ref{gammapoint}.  Incorporation of the MIC integrals into a periodic 
version of the {\sc ONX} algorithm \cite{ESchwegler97} for linear scaling computation of the HF 
exchange matrix is then described in Section \ref{implementation}.  In Section \ref{validation} 
we demonstrate convergence of the HF-MIC model chemistry to the ${\bf k}$-space integration 
limit for a number of systems, and in Section \ref{scaling} we demonstrate linear scaling 
for both diamond and ice.  In Section \ref{discussion} we discuss these results,
and then in Section \ref{conclusions} we summarize our results.


\section{Periodic exact exchange}

In the conventional formulation of periodic boundary conditions, the 
Bloch functions 
\begin{equation}
\psi^{\bf k}_a({\bf r})  =  \sum_{\bf R} e^{i {\bf k}\cdot {\bf R}} \phi_a ({\bf r}-{\bf R}),
\label{Block}
\end{equation}
are often constructed from non-orthogonal functions local to the unit cell. Here, the 
local function
$\phi_a$ is a Gaussian-Type Atomic Orbital (GTAO) centered on atom {\bf A}, while the 
sum on {\bf R} runs over the Bravais lattice vectors defined by integer translates of the lattice  primitives 
 {\bf a}, {\bf b} and {\bf  c}.  Inversely, the vector {\bf k} runs over 
the reciprocal lattice vectors.

To date, rapid computation of the Hartree-Fock exchange interaction demands
the analytic evaluation of two-electron integrals, which is possible when the 
local basis functions are of Cartesian Gaussian type.  
Typically, these functions have the form
\begin{equation}
\phi_a ({\bf r}) = (x-A_x)^{l_a} (y-A_y)^{m_a} (z-A_z)^{n_a}{\large e}^{-\zeta_a ({\bf r}-{\bf A})^2}
\end{equation}
where the triad $\{l_a,m_a,n_a\}$ sets angular symmetry  
and the exponent $\zeta_a$ is chosen to describe a particular length scale. 
Gaussian basis functions are often contracted to approximate 
atomic eigenfunctions.
 
With periodic boundary conditions, the exact Hartree-Fock exchange matrix is 
\cite{RDovesi00,MCausa88}
\begin{equation}
{\bf K} [{\bf k}] = \sum_{{\bf L}} {\bf K} [{\bf L}] e^{i {\bf k}\cdot {\bf L}} \, ,
\label{Kinkspace}
\end{equation}
where
\begin{equation}
K_{ab} [{\bf L}] = - \frac{1}{2}
\sum _{{\bf H N},c d} P_{cd}[{\bf N}]
\left(
      \phi        _a    
      \phi^{\bf H}_c    
{\big | }
      \phi^{\bf L}_b    
      \phi^{\bf H+N}_d  
\right) 
\label{CryEq}
\end{equation}
is defined in real space, and the two-electron integrals, written in  chemist's notation, are 
\begin{eqnarray}
\left(
      \phi        _a  
      \phi^{\bf H}_c  
{\big | }
      \phi^{\bf L}_b  
      \phi^{\bf H+N}_d
\right)
= \quad\quad\quad\quad\quad 
 \quad\quad\quad\quad\quad  
 \quad\quad\quad\quad\quad 
\nonumber \\
\iint d {\bf r} d {\bf r}'
\frac{{\phi_a({\bf r}) \phi_c({\bf r}+{\bf H}})\phi_b({\bf r'}+{\bf L})\phi_d({\bf r'}+{\bf H}+{\bf{N})}}
{{\left|{\bf r}-{\bf r'}\right|}} \, .
\nonumber\\
\end{eqnarray}

The formally infinite sums over lattice vectors in Eq.~(\ref{CryEq}) involve 
many contributions that are in practice infinitesimal.  In part this is due to 
the decay between local basis function products $\phi_a \phi_c $; the product of
two Guassians centered at ${\bf A}$ and ${\bf C}$ decays also as a Gaussian with 
separation $|{\bf A}-{\bf C}|$. 
Truncation based purely on the overlap of Gaussian basis functions is reliable and
well controlled, leading  to ${\cal O}(N)$ product terms.
Additionally, the density matrix is known to decay exponentially 
for non-metallic systems.  A cutoff defining the range of allowed exchange 
interactions may be used to exploit this fall off {\em a priori} by confining summation over 
${\bf N}$ \cite{RDovesi00,MCausa88,REuwema74,CPisani80,RDovesi80} to only those terms that 
satisfy the imposed geometric constraints.  Methods equivalent to a cutoff have also 
been used to achieve $N$-scaling of the exchange matrix in gas phase calculations
\cite{ESchwegler96,JBurant96}.  

\section{Exchange at the $\Gamma$-point}\label{gammapoint}

The $\Gamma$-point approximation limits {\bf k}-space sampling to just the central cell at
${\bf k} = 0$.   In the naive $\Gamma$-point limit, we take ${\bf P[ N ] } \equiv \delta_{\bf N,0} \, {\bf P} $,
reducing Eq.~(\ref{CryEq}) to the less cumbersome relation:
\begin{equation}
K_{ab}= -\frac{1}{2}
\sum _{{\bf H L}, c d} P_{cd}
\left(
      \phi_a    
      \phi^{\bf H}_c     
{\big | }
      \phi_b  
      \phi^{\bf L}_d  
\right)\, ,
\label{CryEq_2}
\end{equation}
where we have relabeled the lattice sums for clarity.
However, there are two significant problems with this naive approach.
First, the exact exchange potential, which is periodic in Eq.~\ref{CryEq}, has been truncated 
asymmetrically to yield Eq.~\ref{CryEq_2}. Neighboring cells correctly representing the exchange interaction  
have been removed,  leading to a non-periodic exchange potential that varies with position of the coordinates in relation 
to the unit cell.   Effectively, a position dependent environment has been created for each charge distribution.
Secondly, this truncated expression violates symmetry of the generalized exchange energy:
\begin{equation}
E_x[{\bf P}^a,{\bf P}^b] \ne E_x[{\bf P}^b,{\bf P}^a] \, , 
\end{equation}
where 
\begin{equation}
E_x[{\bf P}^a,{\bf P}^b] = {\rm Tr} \left[  {\bf P}^a \cdot {\bf K}\left({\bf P}^b\right)  \right] \, ,
\end{equation}
and the superscripts label different density matrices. 

Let us consider the first problem, which is the most serious defect of  Eq.~\ref{CryEq_2}.
This translational variance has a well understood analogue in classical molecular dynamics 
\cite{NMetropolis53,MAllen90,MHloucha98}, where the arbitrary truncation of short range potentials also leads 
to numerical artifacts in energies and forces.  In both cases, translational invariance is  restored by introducing 
the Minimum Image Convention (MIC), which ensures that interactions are always calculated between nearest images.
In computation of the Hartree-Fock exchange matrix, the MIC $\Gamma$-point approximation is just
\begin{equation}
K_{ab}=-\frac{1}{2}
\sum _{{\bf H L}, c d} P_{cd}
\left(
      \phi        _a    
      \phi^{\bf H}_c    
{\big | }
      \phi        _b    
      \phi^{\bf L}_d  
\right)_{\rm  MIC},
\label{MIC}
\end{equation}
where the MIC condition is applied in computation of the two-electron integrals
at the contraction phase, ensuring that primitive charge distributions 
interact consistently over a minimum distance.  In particular, if the primitive basis 
function product $\phi_a \phi^{\bf H}_c$ is centered at ${\bf P}$ and the primitive product 
$\phi_b \phi^{\bf L}_d$ is at ${\bf Q}$, then the minimum image convention is 
applied to the interaction vector ${\bf PQ} \equiv {\bf P}-{\bf Q}$ using
\begin{subequations}
\begin{eqnarray}
{\bf pq}&=&{\bf M}^{-1} \cdot {\bf PQ} \\
pq_i & =& pq_i - {\tt ANINT} \left( pq_i -{\tt SIGN}\left( \delta,pq_i\right) \right) \\
{\bf PQ}_{\rm MIC}&=&{\bf M} \cdot {\bf pq} 
\end{eqnarray}
\end{subequations}
where $\delta \approx 10^{-15}$  is needed to avoid wrapping errors 
\footnote{$\delta$ is required to yield a consistent wrapping when 
distributions lie exactly at the cell boundary.  In effect, this implementation 
changes the wrapping condition from $|pq_i| \ge 1$ to $|pq_i| > 1$.
{\tt ANINT}($x$) rounds $x$ to the closest integer and
{\tt SIGN}($x,y$) transfers the sign from  $y$ to  $x$.
}, 
and ${\bf  M}$ is the $3 \times 3$ shape matrix of the unit cell, 
composed of the primitive lattice vectors,
\begin{equation}
{\bf M} = \left(
\begin{array}{ccc}
a_x & b_x & c_x \\
a_y & b_y & c_y \\
a_z & b_z & c_z 
\end{array} \right)
\end{equation}
This approach is completely general, and can be used at the primitive level with any modern approach
to computing two-electron integrals. While the MIC yields translational invariance of the
exchange matrix, it  does not recover Exchange Kernel Permutational Symmetry (EKPS).  With greater 
expense, we could recover the EKPS by  explicitly symmetrizing the primitive Gaussian 
products within the kernel.  However, in the limit of large systems, where the range of the density 
matrix becomes smaller than the system size, the  EKPS is recovered as demonstrated in Section 
\ref{validation} (note that, without the MIC, EKPS is  not recovered in this limit).

\section{Optimal damping and symmetry of the exchange kernel}

For difficult, unstable SCF problems, the Optimal Damping Algorithm (ODA) of Canc{\'e}s \cite{ECances00} is
an efficient method that guarantees convergence of the HF model.  
However,  permutational symmetry of the exchange kernel is an implicit, simplifying assumption in formulation
of the conventional ODA algorithm.   For small periodic systems, violation of the EKPS creates problems
for the ODA, leading to a non-quadratic behavior (the HF model should yield an exactly quadratic, convex 
minimization problem).   While the EKPS is restored with increasing system size, loose numerical thresholds
can also lead to loss of the EKPS in the limit of a large system.  
In both cases, loss of EKPS can lead to incorrect determination of the ODA mixing parameter
\begin{eqnarray}
\lambda = {{\frac{d E^0}{d \lambda}} \over {3E^1-3E^0-2{\frac{d E^0}{d \lambda}}-{\frac{d E^1}{d \lambda}}}} \, ,
\end{eqnarray}
where the superscripts indicate consecutive steps in the SCF cycle given by the endpoints, 
with $\lambda \in [0,1]$.   It is of course a simple matter to reformulate the ODA, using definitions 
for the endpoint derivatives that do not assume EKPS:
\begin{eqnarray}
\frac{d E^0}{d \lambda} &&=  E^{1}_{\rm ne}-E^{0}_{\rm ne}  
+{\rm Tr}\left[\left( {\bf P}^1 - {\bf P}^0 \right) {\bf T}  \right]\nonumber\\
&& +{\rm Tr}\left[ {\bf P}^1 {\bf F}^0 \right] 
   +{\rm Tr}\left[ {\bf P}^0 {\bf F}^1 \right] 
   -2 {\rm Tr}\left[ {\bf P}^0 {\bf F}^0 \right] \nonumber\\
\frac{d E^1}{d \lambda} &&=  E^{1}_{\rm ne}-E^{0}_{\rm ne}  
+{\rm Tr}\left[\left( {\bf P}^1 - {\bf P}^0 \right) {\bf T}  \right]\nonumber\\
&& -{\rm Tr}\left[ {\bf P}^1 {\bf F}^0 \right] 
   -{\rm Tr}\left[ {\bf P}^0 {\bf F}^1 \right] 
   +2 {\rm Tr}\left[ {\bf P}^1 {\bf F}^1 \right] \, , \nonumber\\
\end{eqnarray}
where  ${\bf F}$ is the Fockian, ${\bf T}$ is the kinetic energy matrix  and $E_{\rm ne}$ is the nuclear-electrostatic energy.
This  modified ODA leads to the correct quadratic parameterization and guarantees convergence of the HF-MIC model.  


\section{Implementation}\label{implementation}

A general treatment of $\Gamma$-point periodic boundary conditions has been implemented in the {\sc MondoSCF}\cite{MondoSCF}
suite of programs for linear scaling quantum chemistry.  A detailed account of these developments for 
pure Density Functional Theory has been given in a companion paper\cite{CTymczak04A}, including the periodic 
development of the Quantum Chemical Tree Code ({\sc QCTC}) for $N$-scaling Coulomb summation.  In addition to QCTC, 
the linear scaling, Quartic Trace-ReSetting ({\sc TRS4}) \cite{ANiklasson03} density matrix solver has been used throughout, 
together with inverse congruence transformations provided by sparse atom-blocked approximate {\sc AINV} \cite{MBenzi01}.  

The Order N eXchange ({\sc ONX}) algorithm \cite{ESchwegler97} for computing the gas phase exchange matrix 
has been modified by placing dual loops running over the lattice vectors 
${\bf H}$ and ${\bf L}$ around the original ONX loop structures.  Two ordered bra and ket distribution 
buffers are assembled for each lattice vector pair, which are then used to drive the basic {\sc ONX} algorithm.
The Minimum Image Convention has been introduced into the primitive contraction stage, in the 
Vertical Recurrence Relations component of a symmetry driven Head-Gordon Pople \cite{MHeadgordon88} scheme for 
computing two-electron integrals.  

All developments  were implemented in  {\sc MondoSCF} v1.0$\alpha$9 \cite{MondoSCF}, a suite of 
linear scaling Quantum Chemistry codes.  The code was compiled using the Portland 
Group F90 compiler {\sc pgf90} v4.2 \cite{pgf90} with the {\tt -O1 -tp athlon} options  and with the 
GNU C compiler {\sc gcc} v3.2.2 using the {\tt -O1} flag.  All calculations were carried out on a 
1.6GHz AMD Athlon running RedHat  {\sc Linux}v9.0 \cite{RedHat90}.   

\begin{table}[ht]
\caption{Progression of Hartree-Fock $\Gamma$-point super-cell calculations 
for MgO using the periodic RHF-MIC and RHF 8-511G/8-51G level of theory.  
Comparison is 
made to a final value approaching the ${\bf k}$-space integration limit for 
the primitive cell.}
\label{MgOTable}
\center{\begin{tabular}{lrll}
\toprule
Program         & $N_{\rm MgO}$              & Energy (au)    & Energy/$N_{\rm MgO}$\\ 
\colrule
{\sc MondoSCF}       & 2$^b$    & -274.530033     & -137.265017 \\
                     & 8$^c$    & -1098.43823     & -137.304778  \\
                     & 16$^b$   & -2197.05636     & -137.316023  \\
                     & 32$^c$   & -4394.48467     & -137.327646  \\
                     & 54$^b$   & -7415.89617     & -137.331411  \\
                     & 64$^c$   & -8789.24941     & -137.332022  \\
                     & 128$^b$  & -17578.5022     & -137.332048  \\
                     & 216$^c$  & -29663.7291     & -137.332079  \\ 
\hline
{\sc CRYSTAL98}$^a$  & 2$^d$    & -274.664153     & -137.332077  \\ 
\botrule 
\end{tabular}\\}
\raggedright{
{\hskip 0.18in}$^a \Gamma$-point\\
{\hskip 0.18in}$^b$Triclinic \\
{\hskip 0.18in}$^c$Cubic \\
{\hskip 0.18in}$^d 8\times8\times8$ ${\bf k}$-space integration grid  \\}
\end{table}

\begin{table}[h]
\caption{Progression of $\Gamma$-point super-cell calculations of proton ordered ice
at the RHF-MIC/8-51G/5-11G$^*$ level of theory.   Comparison is made to a final value 
approaching the ${\bf k}$-space integration limit for the primitive cell.}
\label{PIceTable}
\center{\begin{tabular}{lrccc}
\toprule
Program             & $N_{\rm H_20}$ & Energy (au)    & Energy/$N_{\rm H_2O}$ & DEKPS\\ 
\colrule
{\sc MondoSCF}$^a$  & 2              &  -152.03025  &  -76.01512  & $10^{-3}$\\
                    & 16             &  -1216.3003  &  -76.01877  & $10^{-4}$ \\
                    & 54             &  -4105.0163  &  -76.01882  & $10^{-5}$ \\
                    & 128            &  -9730.4090  &  -76.01882  & $10^{-6}$ \\
                    & 250            &  -19004.705  &  -76.01882  & $10^{-6}$ \\ 
\hline
{\sc CRYSTAL98}$^b$  & 2   &  -152.03765  &  -76.01882  \\ 
\botrule
\end{tabular}\\}
\raggedright{
{\hskip 0.01in}$^a \Gamma$-point\\
{\hskip 0.01in}$^b 6\times6\times6$ ${\bf k}$-space integration grid \\}
\end{table}

\begin{table}[h]
\caption{Lattice constants in \AA~for diamond computed using different system sizes, 
theory levels, and basis sets at a {\tt LOOSE} accuracy. For comparison, 
the experimental value for diamond, extrapolated to $T = 0$K, is 
${\rm a}_0 = 3.567$\AA, while ${\rm a}_0 = 3.583$\AA~is obtained 
with both the RPBE/6-31G* GGA and the RTPSS/6-31G* meta-GGA level of theory \cite{VStaroverov04}.}
\label{DiamondLC}
\center{\begin{tabular}{crcccc}
\toprule
Program              &$N_{\rm at}$   & Basis         & ${\rm a}_0^{\rm \scriptscriptstyle HF}$ & ${\rm a}_0^{\rm \scriptscriptstyle PBE0}$ \\
\colrule
{\sc MondoSCF}$^a$   &~64~           & ~STO-3G~      & ~3.587~          &  ~3.601~    \\
      --             &~216~          & ~STO-3G~      & ~3.582~          &  ~3.594~    \\
{\sc CRYSTAL}$^b$    & ~2~           & ~STO-3G~      & ~3.581~          &   -- \\ 
{\sc MondoSCF}$^a$   &~64~           & ~6-21G*~      & ~3.575~          &  ~3.571~    \\
      --             &~216~          & ~6-21G*~      & ~3.571~          &  ~3.569~    \\
{\sc CRYSTAL}$^b$    & ~2~           & ~6-21G*~      & ~3.574~          &   -- \\
\hline
\botrule
\end{tabular}\\}
\raggedright{
{\hskip 0.33in}$^a \Gamma$-point\\
{\hskip 0.33in}$^b$ Taken from Ref.~[\onlinecite{ROrlando90}]\\}
\end{table}

\begin{figure}
\caption{The relaxed, uniaxial lattice potential of proton ordered ice \cite{SCasassa97}.
Comparison is made between a RHF-MIC/8-51G/5-11G$^*$ 250 molecule $\Gamma$-point super-cell 
calculation and a {\sc CRYSTAL98} calculation carried out with a two molecule primitive
cell using a~$6\times6\times6$ ${\bf k}$-space integration grid.}
\label{IceEnergyVsLattice}
{\center \includegraphics{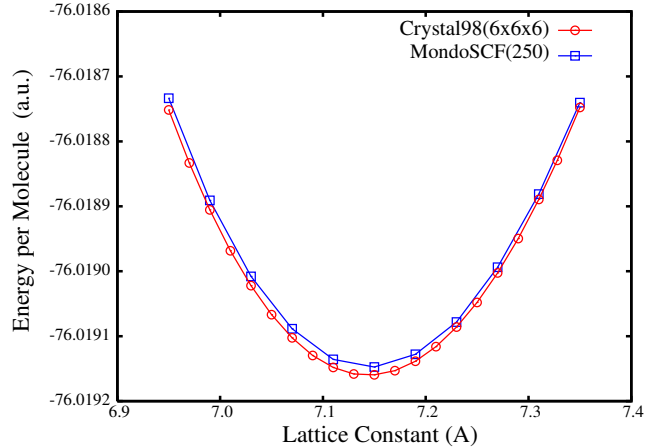}\par}
\end{figure}

\section{Validation} \label{validation}

Comparison is made to the Gaussian orbital, periodic {\em ab initio} program {\sc CRYSTAL98} \cite{CRYSTAL98}, 
primarily using basis sets optimized for the condensed phase, obtained  from Ref.~[\onlinecite{TowlerLib}].

Table \ref{MgOTable} shows the progression of total energies computed with {\sc MondoSCF} 
for MgO with the  $\Gamma$-point RHF-MIC model, using the 8-511G basis set for magnesium and the 
8-51G basis set for oxygen.   Comparison is made to a final value approaching the ${\bf k}$-space 
integration limit for the primitive cell obtained using {\sc CRYSTAL98}.   The {\sc CRYSTAL} basis 
sets were obtained from Ref.~[\onlinecite{TowlerLib}], and the primitive cubic and triclinic cell 
coordinates used for this system are given in Ref.~[\onlinecite{PBCCoordinates}].   The values 
controlling accuracy of the {\sc CRYSTAL98} program were obtained from Ref.~[\onlinecite{BCivalleri02}], 
while the {\sc MondoSCF} calculations were carried out using the {\tt TIGHT} level of accuracy, 
defining numerical thresholds delivering numbers precise to the digits quoted. 

In Table~\ref{PIceTable}, total energies computed with the RHF-MIC $\Gamma$-point
super cell approach are listed for proton ordered ice \cite{SCasassa97}, using the 8-51G basis 
for oxygen and the 5-11G$^*$ basis for hydrogen.  A comparison is made to a final 
value obtained using {\sc Cyrstal98} with a $6\times6\times6$ ${\bf k}$-space integration grid
as explained in Ref.~[\onlinecite{SCasassa97}].  The {\sc MondoSCF} values were obtained 
using the {\tt GOOD} accuracy level, defining numerical thresholds that deliver total energies
precise to the digits quoted.
The primitive cubic cell coordinates used in these ice calculations are listed in Ref.~[\onlinecite{PBCCoordinates}].      
Also shown in this table is a measure of the 
Deviation from the Exchange Kernel Permutational Symmetry (DEKPS),
\begin{equation}
{\rm DEKPS} = || {\bf P}^0 {\bf K}^1-{\bf P}^1 {\bf K}^0 ||_2/|| {\bf P}^1 - {\bf P}^0||_2 \, ,
\end{equation}
where the superscripts refer simply to consecutive steps in the SCF cycle.  Because this measure
is normalized, it yields a roughly consistent measure throughout the SCF.  As shown in Table \ref{PIceTable},
the DEKPS decreases with system size to a constant value consistent with the numerical threshold used to 
discard matrix elements.   

Table~\ref{DiamondLC}  compares the diamond lattice constant computed by MondoSCF using the RHF-MIC 
$\Gamma$-point super-cell approximation to results obtained in reference [\onlinecite{ROrlando90}].  
The result achieved using the 216 atom $3 \times 3 \times 3$ 
supercell, the 6-21G* basis set and the hybrid PBE0 functional yields $a_0=3.569$\AA~and should be
compared to the experimental, $T=0$K result of $a_0=3.567$\AA, as well as 
the value $a_0=3.583$\AA~computed with the TPSS meta-GGA\cite{VStaroverov04}.

Figure~\ref{IceEnergyVsLattice} shows the unrelaxed, uniaxial lattice potential of 
proton ordered ice \cite{SCasassa97} in the $a$ direction (see Ref.~[\onlinecite{PBCCoordinates}] 
for the coordinate system).  Comparison is made between a 250 molecule RHF-MIC $\Gamma$-point 
super-cell calculation performed with {\sc MondoSCF} and {\sc CRYSTAL98} calculations carried 
out with a two molecule primitive cell using a~$6\times6\times6$ ${\bf k}$-space integration grid, 
the 8-51G basis for oxygen and the 5-11G${^*}$ basis for hydrogen.  The {\sc MondoSCF} {\tt GOOD} 
option was used, delivering a relative accuracy of 6-7 digits. The potential minimum for the 
{\sc CRYSTAL98} curve is $7.144$\AA~and  for the {\sc MondoSCF} calculation $7.145$\AA.

\begin{figure}[h]
\caption{CPU time for the exchange, Coulomb  and density 
matrix build (scaled by 10) of diamond using {\tt LOOSE} thresholding 
at the RHF-MIC/STO-3G level of theory.}
\label{DiamondScaling_1}
{\centering \includegraphics{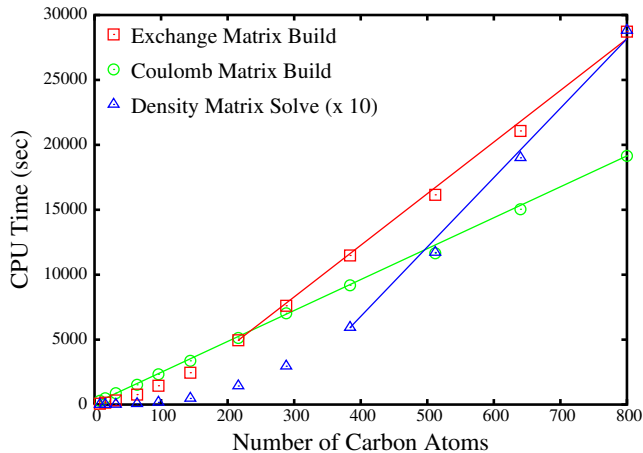} \par}
\end{figure}

\begin{figure}[h]
\caption{CPU time for the exchange, Coulomb and density 
matrix build (scaled by 10) of proton ordered ice using {\tt GOOD}
thresholding at the RHF-MIC/8-51G/5-11G$^*$ level of theory.}
\label{IceScaling}
{\centering \includegraphics{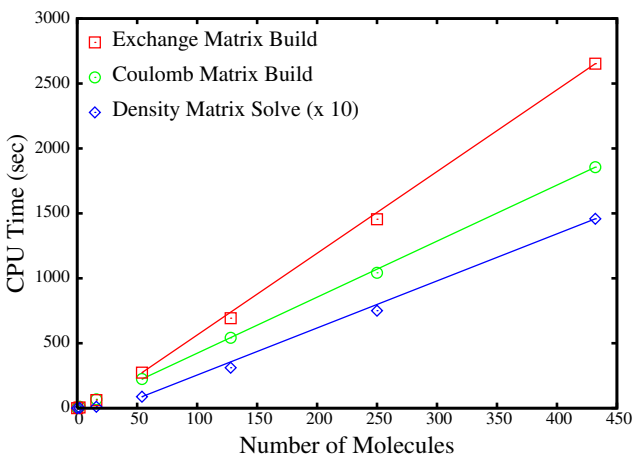} \par} 
\end{figure}

\section{Scaling}\label{scaling}

In Fig.~\ref{DiamondScaling_1}, timings are shown for building the exact exchange, the Coulomb, 
and the density matrix of diamond at the RHF-MIC/STO-3G level of theory with  a {\tt LOOSE} 
set of thresholds.  This is the same accuracy level used to compute the values listed in Table~\ref{DiamondLC}.

Figure~\ref{IceScaling} shows timings for building the  exchange, Coulomb and density matrix
of proton ordered ice \cite{SCasassa97} at the RHF-MIC model, {\tt GOOD} thresholds, the 8-51G basis 
set for oxygen and the 5-11G$^*$ basis set for hydrogen.  This is the level of theory used to
produce Table~\ref{PIceTable} and Fig.~\ref{IceEnergyVsLattice}.  An early onset of linear scaling 
is observed at about 50 water molecules, which is similar to the onset achieved for water clusters 
in the gas phase \cite{ANiklasson03}.

\section{DISCUSSION}\label{discussion}

Differences between {\sc MondoSCF} and {\sc CRYSTAL98} include the use of fitting functions and $\bf k$-space 
integration by {\sc CRYSTAL98} and the use of approximate ${\cal O}(N)$ scaling algorithms and the 
$\Gamma$-point approximation by {\sc MondoSCF}.  Nevertheless, in the large super cell limit, we have achieved a 
very close agreement between the two methods.   

Importantly, the periodic {\sc MondoSCF} algorithms have also 
demonstrated linear scaling for three-dimensional systems.  While 
achieving linear scaling in three-dimensions for gas phase systems is challenging, 
it is even more difficult with full periodicity.  This is especially true for
dense systems like diamond;  linear scaling was achieved here only for large unit cells, 
a minimal basis and a {\tt LOOSE} accuracy level.  It is possible to achieve
an early onset of linear scaling with significantly larger Gaussian basis sets by 
tightening the valence region, but doing so without care can lead to  poor 
performance in the computation of proprieties such as the lattice constant.  
Localization of the  basis set  has received recent, critical attention \cite{SKenny00,JJunquera01,EAnglada02},
leading to numerical  basis functions that yield both linear scaling and high quality properties.  So far though,
this issue remains unexplored in the context of a Gaussian representation.

For less dense systems and larger basis functions, the periodic {\sc MondoSCF} algorithms 
are able to achieve an early onset of linear scaling with more accurate thresholding parameters.
With parallelism \cite{MChallacombe00B,CGan03,CGan04B}, this opens the prospect of performing hybrid HF/DFT simulations of fluids. 

These differences in behavior highlight the use of continuous thresholds that set the cost 
to accuracy ratio on the fly.  Thus, as the gap closes or the basis sets becomes overcomplete,  
periodic {\sc ONX} and {\sc TRS4} will correctly revert to ${\cal O}(N^2)$ and ${\cal O}(N^3)$ algorithms 
respectively.  This should be compared with  use of a radial cutoff to determine
{\em a priori } the  graph of the density and exchange matrix, which does not correctly  
permit fill-in.

\section{CONCLUSIONS}\label{conclusions}

For the first time, a translationally invariant definition of the periodic 
Hartree-Fock $\Gamma$-point approximation has been presented.  Based on inserting 
the Minimum Image Condition (MIC) into the contraction phase of periodically
summed two-electron integral algorithms,   this HF-MIC approach has been used 
to extend the {\sc ONX} algorithm for linear scaling computation of the exchange 
matrix to periodic boundary conditions.  Convergence of the HF-MIC $\Gamma$-point
super-cell approximation to the ${\bf k}$-space integration limit has been 
demonstrated for MgO and ice to better than 8 digits.  Linear scaling was 
demonstrated for diamond and ice, including MIC-exchange, Coulomb and density matrix 
construction.   

\section*{ACKNOWLEDGMENTS}

We would like to acknowledge Tommy Sewell and Ed Kober for their advice
and support. We would also like to thank Chee Kwan Gan and Anders Niklasson 
for a careful reading of this manuscript.  

\bibliographystyle{apsrmp} 
\bibliography{hf} 

\appendix

\section{Thresholds}\label{Thresholds}

There are four thresholds that control the numerical precision of the linear scaling algorithms
in {\sc MondoSCF}.  They are the matrix threshold $\tau_{\rm \scriptscriptstyle MTRIX}$
described in Ref.~[\onlinecite{ANiklasson03}], the two-electron threshold  $\tau_{\rm \scriptscriptstyle 2E}$ ({\tt thresh} in Ref.~[\onlinecite{ESchwegler97}]),
the distribution threshold   $\tau_{\rm \scriptscriptstyle DIST}$ described in Ref.~[\onlinecite{CTymczak04A}],
and the HiCu threshold $\tau_{\rm \scriptscriptstyle HICU}$ used to set the exchange-correlation grid ($\tau_r$ in 
Ref.~[\onlinecite{MChallacombe00A}]).  The threshold $\tau_{\rm \scriptscriptstyle 2E}$ also controls accuracy of the
Coulomb matrix as discussed in Ref.~[\onlinecite{CTymczak04A}].
These thresholds, listed in Table~\ref{TableOfThresholds},  have been 
calibrated over a wide range of systems to yield a minimum of  4, 6 and 8 digits of relative accuracy in the total 
energy with {\tt LOOSE}, {\tt GOOD} and {\tt TIGHT} respectively.    Typically however, one additional digit is achieved for 
models that contain Hartree-Fock exchange.

\begin{table}[h]
\caption{Accuracy goals and corresponding thresholds that control precision of the {\sc MondoSCF} linear scaling algorithms.}
\label{TableOfThresholds}
\center{
\begin{tabular}{lllll}
\toprule
Accuracy        &  $\tau_{\rm \scriptscriptstyle MTRIX} $ & $\tau_{\rm \scriptscriptstyle 2E}$ & $\tau_{\rm \scriptscriptstyle DIST}$ & $\tau_{\rm \scriptscriptstyle HICU}$ \\ 
\colrule
{\tt LOOSE}     & $10^{-4}$ & $10^{-6}$  & $10^{-8}$  & $10^{-3}$ \\
{\tt GOOD}      & $10^{-5}$ & $10^{-8}$  & $10^{-10}$ & $10^{-5}$ \\
{\tt TIGHT}     & $10^{-6}$ & $10^{-10}$ & $10^{-12}$ & $10^{-7}$ \\
\botrule
\end{tabular}}
\end{table}

\end{document}